# Validation of an ECAPA-TDNN system for Forensic Automatic Speaker Recognition under case work conditions

Francesco Sigona[1] & Mirko Grimaldi

*Centro di ricerca interdisciplinare sul linguaggio (CRIL), University of Salento, Lecce, Italy*

**Abstract**

Different variants of a Forensic Automatic Speaker Recognition (FASR) system based on Emphasized Channel Attention, Propagation and Aggregation in Time Delay Neural Network (ECAPA-TDNN) are tested under conditions reflecting those of a real forensic voice comparison case, according to the *forensic_eval_01* evaluation campaign settings. Using this recent neural model as an embedding extraction block, various normalization strategies at the level of embeddings and scores allow us to observe the variations in system performance, in terms of discriminating power, accuracy and precision metrics. From the achieved results it is possible to state that ECAPA-TDNN can be very successfully used as a base component of a FASR system, managing to surpass the previous state of the art, at least in the context of the considered operating conditions.

**1. Introduction**

Forensic Speaker Recognition (FSR)—also known as Forensic Speaker Identification (FSI) and Forensic Voice Comparison (FVC)—is a method used to determine whether a person's voice can be used as evidence in court. In a typical scenario, FSR is employed to recognize an anonymous offender whose voice was recorded during wire or environmental tapping, while the suspect is the individual who is being investigated as the potential offender.

To this purpose, a Bayesian approach has been widely established, the Likelihood Ratio (LR) Framework, recommended by European Network of Forensic Science Institutes (Drygajlo et al., 2015). Accordingly, the task of the forensic scientist is to provide the court with a strength-of-evidence statement in answer to the question: "How much more likely are the observed differences between the known and questioned samples to occur under the hypothesis that the questioned sample has the same origin as the known sample than under the hypothesis that it has a different origin?" (Morrison, 2010). In a typical setting, the different-origin (i.e., the different-speaker, or defense) hypothesis is that the unknown sample originates from a different person selected at random from a relevant population of speakers.

Advancements in digital signal processing and machine learning make it possible to employ fully automatic approaches to the FSR problem. A Forensic Automatic Speaker Recognition (FASR) system is therefore a technical tool that allows the appropriate vocal characteristics to be extrapolated from the voice recordings, and compared with each other, while the human role is almost exclusively restricted to providing the audio recordings and interpreting the output.

FASR usually relies on previous research on Automatic Speaker Recognition (ASR) used in non-forensic settings, therefore the techniques used in FASR, including acoustic analysis, signal processing, modeling, closely resemble those used in ASR for other applications. The main difference in forensic applications is that the final step has to produce a result suitable for presentation and discussion in a forensic case: i.e., in the form of a LR of the two above mentioned alternative hypotheses. Accordingly, a crucial question arises: how to empirically validate FASR systems under casework conditions? In the last years, substantial advancements have been done in this sense, and validation under case work conditions has become a standard practice (cf.

---

[1] Corresponding author: E-mail address: francesco.sigona@unisalento.it

Morrison 2014 for a review, and Morrison et al. 2021 for a consensus on validation of forensic voice comparison; see also Banks et al. 2020). When a FASR system has been empirically validated, it remains to understand whether its output is suitable to be objectively interpreted in court. Recently, a multi-laboratory evaluation of forensic voice comparison systems, under conditions reflecting those of a real forensic case, has been launched: that is, *forensic_eval_01, 2016–2019* (the results are merged in a special issue of the journal *Speech & Communication*). This is an evaluation campaign born from the observation that, in the field of FVC, the need to empirically test the validity and reliability of the system had remained mostly unsatisfied. Between 2016 and 2019, some independent laboratories participated in this evaluation in order to test the principle FASR system available. Testing new approaches on the basis of common experimental conditions allows to get an idea of the performance of different systems in a comparative perspective.

Recently, Emphasized Channel Attention, Propagation and Aggregation in Time Delay Neural Network (ECAPA-TDNN) was presented as a deep-learning model for the Speaker Verification (SV) task, which demonstrated excellent performance compared to the pre-existing state of the art. The aim of this work is to evaluate the performance of FASR systems based on this neural model, under conditions reflecting those of a real FVC casework, according to the previous cited, *forensic_eval_01* setting (Morrison & Enzinger, 2016).

## 2. Materials and methods

The recordings provided in the context of the *forensic_eval_01* setting include a simulated police interview activity (the known-speaker condition) and an information exchange activity via a telephone call from a landline telephone to a call center (the questioned-speaker condition). The original audio files were recorded in soundproof booths at high fidelity. In the information exchange task, each speaker was recorded on a different channel and was in a different audio booth, communicating via a telephone system. In the interview task, the interviewer was face-to-face with the interviewee, but he was relatively far from the microphone and avoided speaking at the same time as the interviewee (for more details on the data collection protocol see Morrison *et al*. 2012). The audio files used for the comparison tests were obtained starting from the high-quality ones with the subsequent application of digital signal processing techniques, to replicate the effects of transmission through telephone systems, with the addition of noise, reverberation, and final compression. The simulated police interview recordings included substantial room reverberation, and a background noise from a ventilation system, while the telephone call recordings included background office noise (multi-speaker babble and typing noises). A detailed description of these procedures is provided in Enzinger *et al*., 2016. Thus, recordings were obtained that reflect both speaking style and recording conditions, both for speakers in the questioned-speaker condition and speakers in the known-speaker condition. The anonymous speaker recordings were truncated to 46 s and the known speaker recordings to 125.694 s (based on the shortest interview length recorded in the database).

The training and test data to be used in the evaluation finally comes from a total of 166 speakers: 88 of whom recorded in three non-simultaneous recording sessions (at intervals of approximately one week), 35 recorded in two non-simultaneous recording sessions, and 44 registered in one session. The training data consists of a total of 423 recordings of 105 speakers (191 recordings in the anonymous speaker condition and 232 in the known speaker condition), while the test data consists of a total of 223 recordings of 61 speakers (61 recordings in condition of anonymous speaker and 162 in condition of known speaker). These sessions allow to perform 111 same-speaker comparisons (from 61 unique speakers) and 9720 different-speakers comparisons (from 3660 unique pairs of speakers).

As part of the evaluation, the FVC software system can be trained with the training data of *forensic_eval_01* and with any other datasets held by the participant. However, the system must be tested, and possibly calibrated, using only the test data provided.

The evaluation metrics chosen to describe the system performance are both numerical and graphical. The numerical metrics taken into consideration were the following (cf. Brümmer and du Preez, 2006; van Leeuwen and Brümmer, 2007; González-Rodríguez et al., 2007; Morrison, 2011; Drygajlo et al., 2015; Meuwly et al., 2016):

- $C_{llr}^{pooled}$ (Log likelihood ratio cost): a single value summarizing the overall system quality, given by the Eq. (1):

$$C_{llr}^{pooled} = \frac{1}{2}\left[\frac{1}{N_{ss}}\sum_{i=1}^{N_{ss}}\log_2\left(1+\frac{1}{LR_{ss_i}}\right) + \frac{1}{N_{ds}}\sum_{j=1}^{N_{ds}}\log_2\left(1+LR_{ds_j}\right)\right] \text{ Eq. (1)}$$

$LR_{ss}$ are the likelihood ratio values in the case of same-speaker comparison, while $LR_{ds}$ are likelihood ratio values in the case of different-speakers comparisons. Since $LR_{ss}$ values much greater than 1 better support the same-speaker hypothesis, and $LR_{ds}$ values much less than 1 better support the different-speakers hypothesis, smaller values of $C_{llr}^{pooled}$ indicate better performances. On the contrary, a system which provides no useful information and always responds with a likelihood ratio of 1 will result in a $C_{llr}^{pooled}$ value of 1.

- 95% CI (95% credible interval): a metric of precision (reliability) of the output of the system. It measures the variability of the resulting multiple likelihood ratio values when a questioned speaker recoding is compared with all available (if any) recordings belonging to a same speaker (which may be the same as the questioned speaker or a different one). This metric will be calculated using the parametric procedure described in Morrison (2011), and is reported on a scale of ± orders of magnitude (= $\log_{10}$ scale).

- $Cllr^{mean}$ (Log likelihood ratio cost, accuracy only): it is a measure of the accuracy (validity) of the output of the system. According to Morrison and Enzinger, 2016, "this is the same as the $C_{llr}^{pooled}$ metric, but whereas all the test results were pooled to calculate $C_{llr}^{pooled}$, for $Cllr^{mean}$ the calculations were performed on the means of the groups defined in the description of the 95% CI metric" (a group being the resulting multiple likelihood ratio values as described above).

- $Cllr^{min}$ (Discrimination loss), a measure of the quality of the extraction stage (see later), i.e., the quality of the score. It is a $C_{llr}$ computed after the LR values from test results have been optimized using the non-parametric pool-adjacent-violators (PAV) procedure, which involves training and testing on the same data. Therefore, this metric it is not representative of the expected performance when new test data are input to the system. According to Meuwly et al. (2016), the discrimination power represents the capability to distinguish amongst forensic comparisons where different propositions are true.

- $Cllr^{cal}$ (Calibration loss): it is equal to the difference $C_{llr}^{pooled}$ - $Cllr^{min}$. It is a measure of the quality of the presentation stage (see later), i.e., of the likelihood ratio calibration.

- EER (Equal Error Rate): another largely used metric to evaluate the discriminating power of the system. Likelihood ratios test values can be combined with some prior odds to achieve posterior odds; in turn, posterior odd can be matched against a threshold to classify a test comparison as same-speaker (prosecutor hypothesis) or different-speakers (defense hypothesis). By this way, false "identification" and false "rejection" error rate can be computed as proportion of wrong classifications. EER is achieved by adjusting the priors and the threshold in such a way that the two error rates are equals, and the resulting error rate is called the EER. In the context of *forensic_eval_01*, EER is calculated using the Receiver Operator Characteristic Convex Hull method (Brümmer and de Villiers, 2013).

Instead, the following graphical metrics were taken into consideration:

- Accuracy and precision metric plot, representing the combination of $Cllr^{mean}$, 95% CI, and $C_{llr}^{pooled}$

- Tippet plot (also see Meuwly, 2000; Morrison, 2010): it superimposes the cumulative proportion of log likelihood ratios with values less than the value on the *x* axis, achieved for same-speaker comparisons, and the cumulative proportion of log likelihood ratios with values greater than the value on the *x* axis, achieved for different-speakers comparisons. In general, further to the right the same-speaker curve and the further to the left the different-speakers curve, the better to the performance of the system. Also, EER can be read off as the y axis value corresponding to the point where the two curves cross.

- Detection Error Tradeoff (DET) plot. This graphic is described in Martin *et al*. (1997), Drygajlo *et al*. (2015), Meuwly *et al*. (2016). It can be related to the Tippet plot, in the sense the DET plot represents a parametric relation between the false "rejection" and false "identification" error rate, in all possible values for priors and threshold values (as described above). In general, the closer the DET curve is to the origin (zero error rates), the better the performance. In the context of *forensic_eval_01*, the plots are drawn using the Receiver Operator Characteristic Convex Hull method (see Brümmer and de Villiers, 2013).

- Empirical Cross Entropy (ECE) plot. The ECE value is calculated by Eq. (2)

$$ECE = \frac{P_{ss}}{N_{ss}}\sum_{i=1}^{N_{ss}} \log_2\left(1 + \frac{1}{LR_{ss_i}\frac{P_{ss}}{P_{ds}}}\right) + \frac{P_{ds}}{N_{ds}}\sum_{j=1}^{N_{ds}} \log_2\left(1 + LR_{ds_j}\frac{P_{ss}}{P_{ds}}\right) \quad \text{Eq. (2)}$$

where $P_{ss}$ and $P_{ds}$ are the priors for same-speaker and different-speakers hypotheses. The plot represents ECE as a function of the prior odds $P_{ss}$ / $P_{ds}$, calculated in three different settings: a) using LR values from the test results, representing the actual system performances: in this case, ECE(0) is equal to the $C_{llr}^{pooled}$; b) using LR values from the test results after the PAV optimization: in this case, ECE(0) is equal to the $C_{llr}^{min}$ ; c) the ECE for a system that always output LR=1. ECE plots can reveal calibration problems. See also Ramos Castro (2007), Ramos and González-Rodríguez (2013), Ramos *et al*. (2013), Drygajlo *et al*. (2015), Meuwly *et al*. (2016).

**2.1. System description up to the score level**

In recent years, Deep Neural Networks (DNN) that map variable-length utterances to fixed-length vectors (generally called *embeddings*), have emerged as state-of-the-art in speaker recognition, supporting, or more often replacing the use of *i-vectors* (Dehak et al., 2011). In a typical non-forensic application, two utterances can be compared by calculating some distance (similarity) metric between the corresponding embedding vectors and comparing the result against a decision threshold. Among different proposal, the x-vector architecture (Snyder *et al*., 2018) and their improvements (Snyder *et al*., 2019; Zeinali *et al*., 2019; Garcia-Romero *et al*., 2020) have shown excellent performances over time. It is a Time Delay Neural Network (TDNN) in which statistic pooling is used to map utterances to embeddings. Based on this architecture, Desplanques *et al*. (2020), recently have proposed the ECAPA-TDNN, that includes multiple enhancements to the baseline TDNN-based x-vector. A detailed description of the proposed architecture and components is out of the scope of the present work. However, it may be convenient to point out that, according to Desplanques *et al*. (2020), enhancements include restructuring the frame layers into 1-dimensional Res2Net modules (Gao *et al*., 2019) with impactful skip connections, also introducing Squeeze-and-Excitation (SE) blocks in these modules to explicitly model channel interdependencies. Moreover, hierarchical features from different network layers are aggregated and propagated, and the statistics pooling module has been improved with channel-dependent frame attention, to make able the network to focus on different subsets of frames during each channel's statistics estimation. Figure 1 represents the schematics of the network topology, which also benefits from the introduction of elements of the popular Residual Network architecture (ResNet , He et al., 2016).

To the present work, we used the ECAPA-TDNN model available on the Hugging Face platform[2], which was pre-trained using Voxceleb 1 + Voxceleb2 data sets and Additive Margin Softmax Loss, already implemented within the SpeechBrain toolkit (Ravanelli *et al.*, 2021), a well-known Open-Source Conversational AI Toolkit.

At frame level input, the model is configured to work with 80 log Mel filter banks energies: indeed, for the x-vector systems, such features have been found to be more effective (see Landini et al., 2020; Alam et al., 2020, Lee et al., 2020) than traditional mel-frequency cepstral coefficients (MFCCs, by Davis et al., 1980). Moreover, the system has 1024 channels in the convolutional layers, the dimension of the bottleneck in the SE-Block and attention module set to 128, the scale dimension *s* in the Res2Block (Gao *et al.*, 2019) is set to 8, and there are 192 nodes in the final fully-connected layer. The model has more than 20M trainable parameters.

The trained network is essentially used to extract a single 192-dimension embeddings vector from every single utterance.

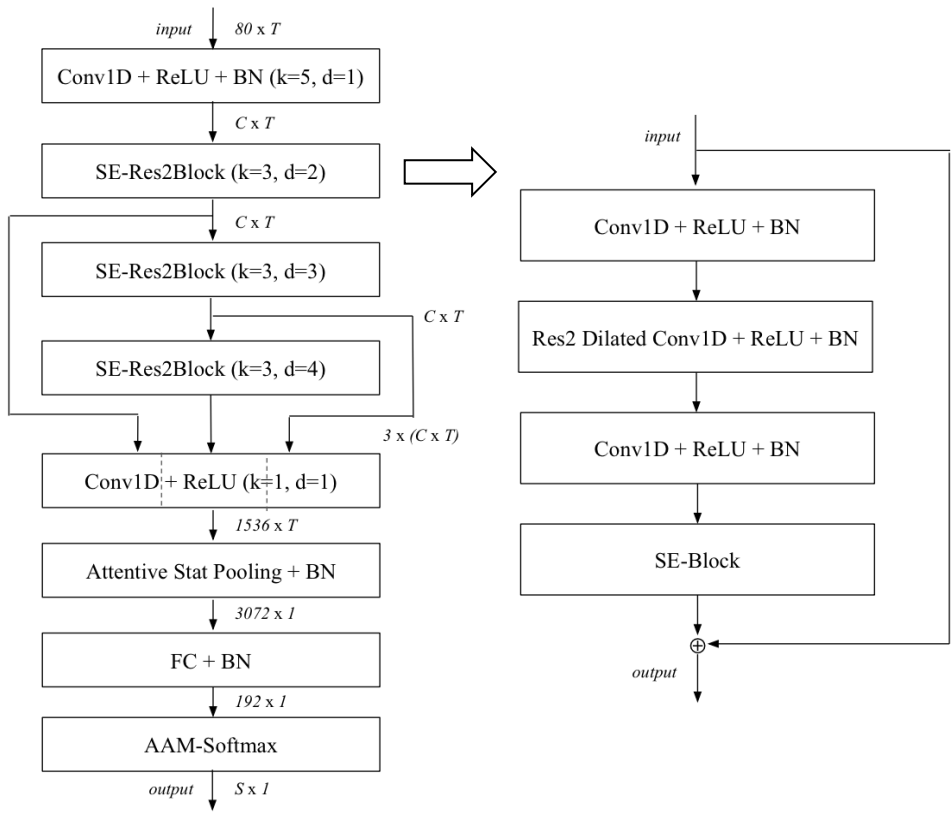

*Figure 1. Left diagram: network topology of the ECAPA-TDNN. k is the kernel size and d is the dilation spacing of the Conv1D layers or SE-Res2Blocks. C (=1024) and T are the channel and temporal dimension of the intermediate feature-maps respectively. S is the number of training speakers. BN stands for Batch Normalization (Ioffe and Szegedy, 2015). Right diagram: the SE-Res2Block of the ECAPA-TDNN architecture. The standard Conv1D layers have a kernel size of 1. The central Res2Net Conv1D with scale dimension s = 8 expands the temporal context through kernel size k and dilation spacing d. Adapted from Desplanques et al. (2020)*

**2.2. Score-level methods and evaluation layout**

In general, a FSR system operation can be decomposed in two sequential stages (Brümmer and du Preez, 2006): the extraction stage, and the presentation stage. The extraction stage refers to the task of extracting information from the input speech into the real score *s*, with the only requirement that more positive scores favour the same-speaker hypothesis, and more negative scores favour the different-speaker hypothesis. The presentation stage refers to presenting the score in a form that is useful to the user application, i.e., in the case of forensic application, in a forensic LR.

---

[2] https://huggingface.co/speechbrain/spkrec-ecapa-voxceleb

In the present work, four system variants have been implemented and tested, all sharing the same architecture and configuration of the ECAPA-TDNN module, which is used to convert each single utterance into the corresponding embeddings vector. This is a first step of the extraction stage. The four systems differ in the second step of the extraction stage, in which the two embeddings to be compared are input to a score computation scheme, to output a single real score that matches the above requirement.

Furthermore, in all cases the presentation stage remains the same: i.e., converting the score to the required forensic LR. This procedure (known as *calibration*) is implemented in the present work by means of the "one-or-two-speakers-leave-out" cross-validation approach, as follows. First, the set of all scores produced by comparing each utterance of the test set in the questioned speaker condition against any other utterance of the test set in the known-speaker condition is calculated. Then, for each score $s_i$ in that set, the LR value for $s_i$ is calculated as the ratio between the same-speaker and the different-speakers probability densities, evaluated in $s_i$. The same-speaker probability density is computed as the Kernel Density Estimate (Parzen, 1962) built on all the same speaker scores except those produced by the speaker (in case $s_i$ was a same-speaker score) or the two speakers (in case $s_i$ was a different-speaker score) that produced $s_i$. The different-speakers probability density is the Kernel Density Estimate built on all the different-speakers scores produced by the same utterance that produced $s_i$. The procedure is repeated for each $s_i$ score of the set. In such a way, there is no chance that the LRs computations were biased due to any information about the test speaker(s).

In a first system variant (SYS1), the comparison score is simply computed as the well-known cosine similarity scoring function between the two input embeddings vectors $\boldsymbol{w_1}$ and $\boldsymbol{w_2}$:

$$score(\boldsymbol{w_1}, \boldsymbol{w_2}) = \frac{\boldsymbol{w_1} \cdot \boldsymbol{w_2}}{\|\boldsymbol{w_1}\|\|\boldsymbol{w_2}\|}$$

while the provided training data are not used at all. This system variant is presented essentially as a baseline for those that follow below.

In a second system variant (SYS2), all the embeddings computed from the entire training data set are used as normalization cohort to implement a Symmetric Normalization (S-norm) (Shum et al., 2010) of the scores. In this procedure, the questioned-condition embedding is compared with each embeddings vector of the cohort, obtaining a set of scores, from which the normalization statistics (mean and standard deviation) are calculated. The suspect condition embedding is also compared with each embeddings vector of the cohort, obtaining a second set of scores, and a second pair of normalization statistics (a second mean and a second standard deviation). Finally, the final normalized score is obtained by normalizing the score with both the statistics sets (subtracting the mean and dividing by the standard deviation), and then taking the average of the two resulting values.

In a third system variant (SYS3), the test embeddings are directly normalized but the scores. All the embeddings computed from the entire training set is considered as normalization cohort and the mean and the standard deviation of each *i-th* component of the embeddings vectors of the cohort are calculated. The *i-th* normalization statistics are then used to normalize the *i-th* component of the two test embeddings. Finally, the cosine similarity score is computed between the two normalized embeddings vectors.

The last system variant (SYS4) is like SYS3 with the difference that the cohort used to normalize the embeddings no longer comes from the entire training set but is automatically selected within the training set. To normalize the questioned-condition embeddings, the cohort composed of the 100 utterances/embeddings in the training set having the highest cosine similarity score is used. Similarly, we proceed for the suspect condition embeddings.

## 3. Results

The results achieved by the four systems (SYS1-4) are represented in Figure 2-3 and in the Table 1. Figure 2 provides Tipplet plots with precision information and ECE plots, while Figure 3 provides plot for DET and combination of Cllr$^{mean}$, 95% CI, and C$_{llr}^{pooled}$ .

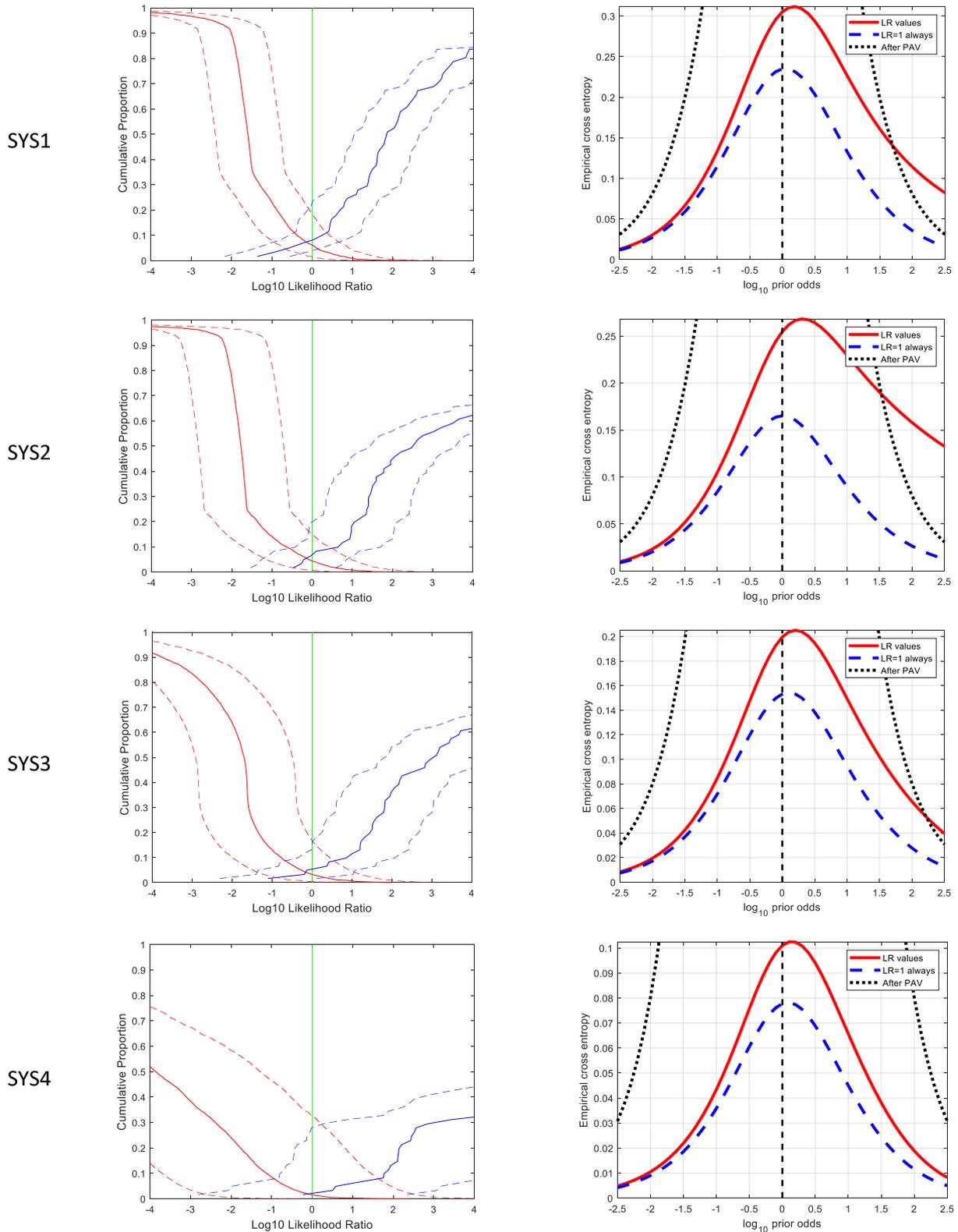

*Figure 2. Graphs on the left of the figure: Tipplet plots with precision. The solid curve that tends to increase represents the cumulative proportion of Log$_{10}$ (LR) with values less than the value on the horizontal axis, achieved for same-speaker comparisons, while the solid curve with tends to decrease represents the cumulative proportion of Log$_{10}$ (LR) with values greater than the value on the horizontal*

axis, achieved for different-speakers comparisons. Dashed lines represent the 95% CI in both cases. Graphs on the right of the figure: ECE plots. The solid line is achieved using LR values from the test results, and it represents the actual system performances: on this curve, ECE(0) is equal to the $Cllr^{pooled}$. The dashed line is achieved using LR values from the test results after the PAV optimization, and ECE(0) is equal to the $Cllr^{min}$. Finally, the dotted line represents a system that always output LR=1, as a reference.

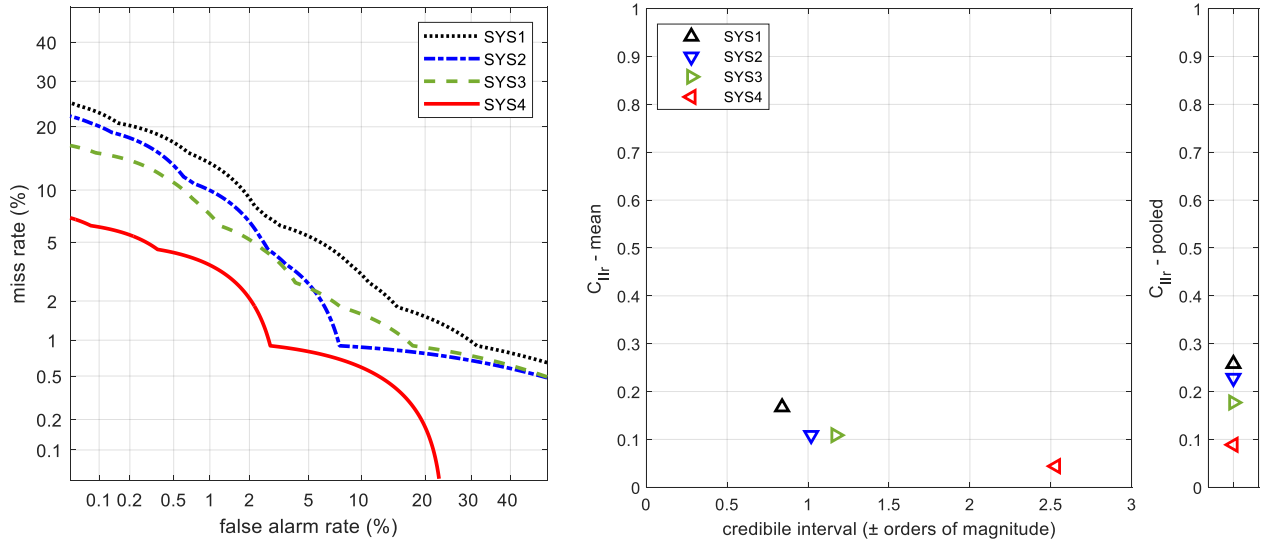

Figure 3. DET plot (first axes on the left), and combination of $Cllr^{mean}$, 95% CI, and $Cllr^{pooled}$, for SYS1-4 variants.

Table 1 reports the numerical metrics achieved by the different variants of the proposed system (SYS1-4) together with the summary outcomes of the *forensic_eval_01* (Morrison and Enzinger, 2019), and the new alpha version of the $E^3FS^3$ ($E^3$ Forensic Science System). The latter is a recent FVC system based on x-vectors and ResNet, being developed by the Forensic Data Science Laboratory at Aston University, with the contributions from other research laboratories (Weber et al. 2022). The various systems are listed in descending order with respect to the value of $Cllr^{pooled}$, so that the systems listed lower in the table performed better overall than the systems that are listed higher.

| System | Type | $Cllr^{pooled}$ | $Cllr^{mean}$ | 95% CI | $Cllr^{min}$ | $Cll^{cal}$ | EER% |
|---|---|---|---|---|---|---|---|
| Batvox 3.1 | GMM-UBM | 0.593 | 0.473 | 1.130 | 0.396 | 0.198 | 12.6% |
| MSR GMM-UBM | GMM-UBM | 0.576 | 0.549 | 0.368 | 0.444 | 0.132 | 13.9% |
| MSR GMM i-vector | GMM i-vector | 0.449 | 0.437 | 0.479 | 0.301 | 0.148 | 8.5% |
| Batvox 4.1 | GMM i-vector | 0.365 | 0.304 | 1.156 | 0.317 | 0.048 | 9.6% |
| Phonexia XL3 | DNN bottleneck | 0.294 | 0.225 | 1.160 | 0.231 | 0.063 | 6.6% |
| Nuance 9.2 | GMM i-vector | 0.285 | 0.258 | 0.336 | 0.161 | 0.124 | 4.7% |
| VOCALISE 2017B | GMM i-vector | 0.267 | 0.230 | 1.178 | 0.239 | 0.029 | 7.0% |
| **SYS1** | **ECAPA-TDNN** | **0.258** | **0.167** | **0.840** | **0.189** | **0.069** | **5.3%** |
| Nuance 11.1 | DNN senone | 0.255 | 0.234 | 0.309 | 0.124 | 0.130 | 3.1% |
| VOCALISE 2019A | x-vector | 0.246 | 0.213 | 1.040 | 0.189 | 0.057 | 5.3% |
| **SYS2** | **ECAPA-TDNN** | **0.228** | **0.109** | **1.019** | **0.131** | **0.097** | **3.6%** |
| $E^3FS^3\alpha$ | x-vector | 0.208 | n.a. | n.a. | n.a. | n.a. | n.a. |
| Phonexia BETA4 | x-vector | 0.208 | 0.163 | 0.779 | 0.098 | 0.110 | 2.2% |
| **SYS3** | **ECAPA-TDNN** | **0.177** | **0.109** | **1.170** | **0.134** | **0.043** | **3.5%** |
| **SYS4** | **ECAPA-TDNN** | **0.089** | **0.044** | **2.531** | **0.065** | **0.024** | **2.0%** |

Table 1. Normal font: performance metrics for the best-performing variant of each system of each system that participated in the forensic_eval_01 (adapted from Morrison and Enzinger, 2019), with the addition of the $E^3FS^3\alpha$ software Weber et al. 2022. Bold font: the performances of the SYS1-4 experimental systems based on ECAPA-TDNN, tested in the present work. In all cases, the reported

*metrics indicate better performance the lower the values. The systems are sorted by decreasing values of Cllr$^{pooled}$, therefore those listed at the bottom of the table perform better overall, based on the experimental settings.*

## 4. Discussion

SYS1, where the comparison score is given by the cosine similarity of the test embeddings, without any kind of normalization, and without using the provided training data, can be considered as a baseline system for the other variants (SYS2-4). Experimental results show that SYS1 cannot compete with either the more recent commercial products such as Nuance 11.1, Vocalize 2019A and Phonexia BETA4, or E$^3$FS$^3$α. This may be due precisely to the fact that the tested configuration does not take advantage of the available training data, which instead are used in the systems just mentioned.

On the other hand, results change considerably for SYS2-4 variants, which instead exploit the training data.

Looking at SYS2, where the training data are used as normalization cohort to perform S-norm score normalization, almost all the metrics improve, up to the level of Vocalise 2019A and beyond. The reason for this improvement probably lies precisely in the introduction of the normalization of the score: this procedure is considered very important in order to improve the performance of the SV task in mismatched conditions, as in the casework of the present evaluation, where one voice is recorded on one type of channel (telephone) and the other on another type (environmental) (on this topic, see, for instance, Ortega-Garcia *et al.*, 1999; Matejka et al., 2017).

With SYS3, all metrics improve compared to SYS2, except 95% CI, which is slightly higher (1.170, compared to 1.019). The difference with SYS2 was that the training data was used not to normalize the scores, but to directly normalize the vectors of embeddings. The strategy seems to work, even if it seems to lose some precision in terms of 95% CI. Also, this variant performed better than E$^3$FS$^3$α and Phonexia BETA4 overall, except for 95% CI, Cllr min, and EER.

Finally, by restricting the cohort for the normalization of the embeddings vectors to the 100 embeddings most similar respectively to the two test utterances/embeddings, SYS4 obtains a further important gain on all metrics, not only compared to SYS3 but also compared to E$^3$FS$^3$α and Phonexia BETA4, for exception, again, of 95% CI.

## 5. Conclusion

This experiment has shown how the recent ECAPA-TDNN model can be very successfully used as a component of a FASR system. The different variants tested were validated on the basis of a typical case, according to the setting of the *forensic_eval_01* evaluation campaign. We showed how strategic the choice of standardization methodology is at the level of scores or embeddings.

Since some commercial tools were also validated within this campaign, it was also possible to compare the performances obtained with ECAPA-TDNN with those systems (although this is not the main purpose of our research).

Furthermore, as rightly pointed out by the promoters of the *forensic_eval_01* campaign, the test-bed concerns a typical but specific case of vocal comparison. However, the results relating to the performance of each comparison system, obtained with the data made available to perform the test, cannot be generalized to all relevant populations and conditions: this because the populations and operating conditions of the comparison can vary greatly from one case to another. In the context of a real case, indeed, the trier of facts should verify whether the forensic voice matching system used in that specific case has been subject to empirical tests of its validity and reliability, using data representative of the relevant reference population and the conditions specific to that case.

For this reason, further tests are underway to verify system performance in different operating conditions (such as duration and number of utterances: Vitolo, 2022; Sigona et al., 2023). Starting from the findings here obtained, our future objective is also to try to improve the system, especially with respect to the precision metrics.

**Funding**

This research did not receive any specific grant from funding agencies in the public, commercial, or not-for-profit sectors.

**Acknowledgments**

The authors acknowledge Dr. Geoffrey Stewart Morrison for making this work possible by providing the datasets and software necessary for the calculation of the *forensic_eval_01* evaluation metrics.